\documentclass[twoside]{dis07}
\usepackage[latin1]{inputenc}
\usepackage[dvips]{graphicx,epsfig,color}
\usepackage{wrapfig,rotating}
\usepackage{amssymb,amsmath,array}

\begin{document}
\title{New Resonances and Meson Spectroscopy at BaBar and Belle}

\author{Vincent Poireau
%
%
\vspace{.3cm}\\
%
Laboratoire de Physique des Particules \\
F-74941 Annecy-le-Vieux - France
}

\maketitle

\begin{abstract}
We present a short review on the recent progresses that have been
made in meson spectroscopy. We discuss the experimental
discoveries made at the BaBar and Belle experiments, as well as
the possible interpretations of the new resonances.
\end{abstract}


\section{Introduction}

Observation of a long list of new meson resonances has been
recently reported by the BaBar and Belle experiments. We present
here the new resonances observed in the $c\bar{s}$ and $c\bar{c}$
sectors.

Analyzes presented here were performed using data collected at the
$\Upsilon(4S)$ resonance with the BaBar and Belle
detectors~\cite{ref:babarBelle}, located at the PEP-II and KEKB
asymmetric energy $e^+e^-$ colliders.

\section{$c\bar{s}$ mesons}

Before 2003, only four $c\bar{s}$ mesons were known: two S-wave
mesons, $D_s$ ($J^P = 0^-)$ and $D_s^*~(1^-)$, and two P-wave
mesons, $D_{s1}(2536)~(1^+)$ and $D_{s2}(2573)~(2^+)$. The masses
predicted by the potential model~\cite{ref:potential} were in good
agreement with the measured masses. The potential model predicted
also two other broad states (width of a few hundred of MeV) at
masses in-between $2.4-2.6$ GeV/$c^2$.

\subsection{$D_{s0}^*(2317)$ and $D_{s1}(2460)$ mesons}

In 2003, two new resonances were discovered by the BaBar and CLEO
experiments: the $D_{s0}^*(2317)$ and $D_{s1}(2460)$
mesons~\cite{ref:firstDsJ}. These two resonances are very narrow,
and have masses well below what was predicted by the potential
model. These states are very well known experimentally: masses are
measured with an error below 2 MeV/$c^2$, 95\% confidence level
upper limits on widths are about 4 MeV~; $J^P$ quantum numbers
($0^+$ and $1^+$ for $D_{s0}^*(2317)$ and $D_{s1}(2460)$
respectively), decay modes and branching fractions are also well
measured. Despite a good knowledge of these states, their
theoretical interpretation is still unclear. One obvious
possibility is to identify these two resonances with the $0^+$ and
$1^+$ $c\bar{s}$ states, although it is difficult to fit these
resonances within the potential model. Other interpretations have
been proposed: four quark states, $DK$ molecules or $D\pi$
atoms~\cite{ref:firstDsJModels}.

\subsection{$D_{sJ}^*(2860)$ meson}

The $D_{sJ}^*(2860)$ resonance was discovered by BaBar in
2006~\cite{ref:DsJAntimo}, looking in $c\bar{c}$ continuum:
$e^+e^- \to D^0 K^+ X$ and $e^+e^- \to D^+ K^0_s X$, where $X$
could be anything. A clear peak is observed in the $DK$ invariant
mass, with a mass of $(2856.6 \pm 1.5 \pm 5.0)$ MeV/$c^2$ and a
width of $(47 ± 7 ± 10)$ MeV. Given that this resonance decays to
two pseudoscalars, the $J^P$ quantum number should be $0^+$,
$1^-$, $2^+$, etc. Different interpretations have been proposed,
inside the $c\bar{s}$ scheme: this state could be a radial
excitation of the $D_{s0}^*(2317)$, but other possibilities are
not ruled out~\cite{ref:DsJAntimoModels}.

\subsection{$D_{sJ}(2700)$ meson}

In the same analysis, BaBar reported a broad enhancement, named
$X(2690)$, at a mass of $(2688 \pm 4 \pm 3)$ MeV/$c^2$ and a width
of $(112 \pm 7 \pm 36)$ MeV. A new state, the $D_{sJ}(2700)$, was
reported independently by Belle at a similar mass, looking at $B^+
\to \bar{D}^0 D^0 K^+$ events~\cite{ref:DsJ2700}. The study of the
$D^0K^+$ invariant mass reveals a clear resonance at a mass of
$(2715 \pm 11 {}^{+11}_{-14})$ MeV/$c^2$ with an internal width of
$(115 \pm 20 {}^{+36}_{-32})$ MeV. An helicity analysis shows that
the favored $J^P$ quantum number is $1^-$. Since the $X(2690)$ and
$D_{sJ}(2700)$ mesons have the same decay modes and that the mass
and width are consistent with each other, it is reasonable to
think that they are indeed the same state.

BaBar did a similar analysis~\cite{ref:DsJ2700}, looking at events
where $B$ decays to $\bar{D}^{(*)} D^{(*)} K$. Thanks to the many
final states studied, this analysis has the advantage to be able
to look at four $D^0K^+$ invariant mass distributions as well as
four $D^+K^0_s$ invariant mass distributions. Adding these final
states together, a clear resonant enhancement is seen around a
mass of 2700 MeV/$c^2$. Also, adding the four $D^{*0}K^+$ and four
$D^{*+}K^0_s$ invariant mass distributions together, a similar
enhancement is observed around a mass of 2700 MeV/$c^2$. No
precise measurement was given by this preliminary analysis yet.

The potential model predicts the $2^3S_1$ $c\bar{s}$ state at a
mass of 2720 MeV/$c^2$. Also, from chiral symmetry considerations,
a $1^+-1^-$ doublet of states has been predicted. If the $1^+$
state is identified as the $D_{s1}(2536)$, the mass predicted for
the $1^-$ state is $2721 \pm 10$
MeV/$c^2$~\cite{ref:DsJ2700Models}.

\section{$c\bar{c}$ mesons}

\subsection{$X(3940)$, $Y(3940)$ and $Z(3930)$ mesons}

Three new states were discovered by Belle at masses around 3940
MeV/$c^2$~\cite{ref:XYZ}. Although their mass are very close to
each other, these new states are thought to be different
resonances. The $X(3940)$ state was discovered in $e^+e^- \to
J/\Psi X(3940)$, looking at the recoiling mass to the $J/\Psi$.
The parameters of this resonance are $M = (3943 \pm 6 \pm 6)$
MeV/$c^2$ and $\Gamma = (15.4 \pm 10.1)$ MeV. This new state was
also seen decaying to $DD^*$, but not $DD$. One possible
interpretation is to identify this resonance with the unobserved
$c\bar{c}$ charmonium state $\eta_c(3S) [3^1S_0]$, although other
interpretations have also been proposed.

A near threshold enhancement was observed by Belle in $B \to
J/\Psi \omega K$, looking at the $J/\Psi \omega$ invariant mass.
This resonance, called $Y(3940)$, has a mass of $(3943 \pm 11 \pm
13)$ MeV/$c^2$ and a width of $(87 \pm 22 \pm 26)$ MeV. This state
could be interpreted as the $c\bar{c}$ state $\chi'_{c1}
[2^3P_1]$.

Finally, a new resonance, the $Z(3930)$, was discovered in $\gamma
\gamma \to D\bar{D}$ with a mass of $(3929 \pm 5 \pm 2)$ MeV/$c^2$
and a width of $(29 \pm 10 \pm 2)$ MeV. One possibility is to
identify this resonance with the $c\bar{c}$ state $\chi'_{c2}
[2^3P_2]$.

\subsection{$X(3872)$ meson}

The $X(3872)$ meson was discovered by
Belle~\cite{ref:X3872Discovery} in $B^\pm \to X(3872) K^\pm$ with
$X(3872) \to J/\psi \pi^+ \pi^-$ in 2003, and quickly confirmed by
the BaBar~\cite{ref:X3872Discovery}, CDF and D0 experiments. Its
mass is known very precisely, $3871.81 \pm 0.36$ MeV/$c^2$, and
its width is less than 2.3 MeV at 90\% confidence level. This
state was also observed in the final state $J/\psi
\gamma$~\cite{ref:X3872OtherStudies}, which implies that its $C$
quantum number is equal to $+1$. The study of the $\pi^+ \pi^-$
invariant mass distribution by Belle and an angular analysis by
CDF shows that $J^{PC} = 1^{++}$ is favored (although $2^{++}$ is
still possible). It has also to be noted that a search for a
charged partner was performed by BaBar, but no signal was
found~\cite{ref:X3872OtherStudies}.

The Belle experiment did a study of the channel $B \to \bar{D}^0
D^0 \pi^0 K$ and observed a clear excess in the $\bar{D}^0 D^0
\pi^0$ invariant mass~\cite{ref:X3872D0Ds0}. The surprise came
from the measure of the mass: $3875.4 \pm 0.7 {}^{+1.2}_{-2.0}$
MeV/$c^2$, which is in disagreement with the mass measured in the
$X(3872) \to J/\psi \pi^+ \pi^-$ channel. This discrepancy was
confirmed by the BaBar experiment~\cite{ref:X3872D0Ds0}, looking
at the $B \to \bar{D}^0 D^{*0} K$ channel (where both decays of
$D^{*0}$, $D^0 \pi^0$ and $D^0 \gamma$, are taken into account).
An excess is observed in the $\bar{D}^0 D^{*0}$ invariant mass,
with a mass of $3875.6 \pm 0.7 {}^{+1.4}_{-1.5}$ MeV/$c^2$. The
masses between Belle and BaBar are in good agreement and are $2.2
\sigma$ away from the $X(3872)$ mass in the $J/\psi \pi^+ \pi^-$
channel. If this excess is due to the $X(3872)$ resonance, then
the quantum number $J^P = 2^+$ is disfavored.

The interpretation of the $X(3872)$ state is rather
difficult~\cite{ref:X3872Models} since there is no satisfactory
$c\bar{c}$ assignment for this resonance.
The coincidence between this resonance mass and the $\bar{D}^0
D^{*0}$ mass led some authors to propose that the $X(3872)$ is a
bound state of the $\bar{D}^0$ and $D^{*0}$ mesons with small
binding energy. One of the prediction of this model is that $B^0
\to X(3872) K^0$ is suppressed by approximately a factor 10
compared to $B^+ \to X(3872) K^+$. Experimentally, this ratio is
measured to $0.50 \pm 0.30 \pm 0.05$ in the $X(3872) \to J/\psi
\pi^+ \pi^-$ channel and to $2.23 \pm 0.93 \pm 0.55$ in the $B \to
\bar{D}^0 D^{*0} K$ channel. It has also been proposed that the
X(3872) resonance is a four quark state. In this case, the model
predicts two neutral states and two charged states, with a
difference of mass between the two neutral states (produced
respectively in $B^0$ and $B^+$ decays) of $(7 \pm 2)$ MeV/$c^2$.
The experimental results show a mass difference of $(2.7 \pm 1.3
\pm 0.2)$ MeV/$c^2$ in the $X(3872) \to J/\psi \pi^+ \pi^-$
channel and $(0.2 \pm 1.6)$ MeV/$c^2$ in the $B \to \bar{D}^0
D^{*0} K$ channel. Other possibilities have been mentioned like
glueball or hybrid state.

\subsection{$Y(4260)$ meson}

The $Y(4260)$ state constitutes also quite a mystery. This new
state, with $J^{PC} = 1^{--}$, was discovered by BaBar in $e^+e^-
\to \gamma_{ISR}(J/\psi \pi^+ \pi^-)$, with a photon radiated in
the initial state~\cite{ref:Y4260}. This resonance was confirmed
by Belle~\cite{ref:Y4260} and CLEO, although masses disagree
between experiments. BaBar measures $M = (4259 \pm 8)$ MeV/$c^2$
and $\Gamma = (88 \pm 23)$ MeV, Belle measures $M = (4295 \pm 10
{}^{+10}_{-3})$ MeV/$c^2$ and $\Gamma = (133 {}^{+26}_{-22}
{}^{+13}_{-6})$ MeV while CLEO measures $M = (4283 {}^{+17}_{-16}
\pm 4)$ MeV/$c^2$. A $3\sigma$ enhancement was also reported by
BaBar in $B \to Y(4260) K^-$, followed by $Y(4260) \to J/\psi
\pi^+ \pi^-$~\cite{ref:Y4260}, although this result needs
confirmation by other experiments. Searches for this resonance
were performed in other channels ($e^+e^- \to \gamma_{ISR} (D
\bar{D}),~e^+e^- \to \gamma_{ISR} (\Phi \pi^+ \pi^-),~e^+e^- \to
\gamma_{ISR} (p \bar{p}),~e^+e^- \to \gamma_{ISR} (J/\psi \gamma
\gamma)$), but no positive results were reported~\cite{ref:Y4260}.

One of the surprise concerning this resonance came from the search
of the $Y(4260)$ going to the decay mode $\psi(2S) \pi^+ \pi^-$ in
ISR production~\cite{ref:Y4325}. A clear signal is observed in
this channel, however with a mass measurement incompatible with
the previous BaBar result. The mass found in this channel is
$(4234 \pm 24)$ MeV/$c^2$ with a width of $(172 \pm 33)$ MeV. This
measurement, although incompatible with the BaBar measurement in
the $J/\psi \pi^+ \pi^-$ channel, is compatible with the Belle
measurement. More data and experiments looking at this channel are
needed to be able to conclude if this excess is due to the
$Y(4260)$.

The interpretation of this state is far from
obvious~\cite{ref:Y4260Models}. There is no $c\bar{c}$ assignment
for a $1^{--}$ state of this mass. This is also probably not a
glueball, since in this case we would have expected a decay to
$\Phi \pi^+ \pi^-$, which was not observed. Other possibilities
are four quark state $[cs][\bar{c}\bar{s}]$, hybrid meson or
$\omega \chi_{c1}$ molecule.

\section{Conclusion}

Although no new resonances were discovered in many years, BaBar
and Belle gave an impressive list of new results since 1999. In
the $c\bar{s}$ sector, the $D_{s0}^*(2317)$ and $D_{s1}(2460)$
mesons are now very well known experimentally, but no definite
interpretation was given theoretically. The $D_{sJ}^*(2860)$ and
$D_{sJ}(2700)$ mesons were discovered recently and need more
experimental inputs. In the $c\bar{c}$ sector, it seems plausible
to identify the $X(3940)$, $Y(3940)$ and $Z(3930)$ mesons to
charmonium states, although other explanations have been proposed.
The $X(3872)$ and $Y(4260)$ resonances are not charmonium states,
and thus are probably the first occurrences of non standard quark
content.

A lot of analyzes are still in progress with the current data set
in BaBar and Belle: more decay modes for the resonances presented
here are being investigated. These two experiments are taking data
until the end of 2008, which is the promise of more surprises to
arise.

The author is very grateful to the organizers of the DIS 2007
conference for their support and all efforts in making this venue
successful.


\begin{footnotesize}


\end{footnotesize}


\end{document}